\documentclass[aip, reprint]{revtex4-2}
\UseRawInputEncoding 
\usepackage{graphicx}
\usepackage{dcolumn}
\usepackage{bm}
\usepackage{sidecap}
\usepackage{xcolor} 
\draft 

\begin{document}

\title{Tailoring of Plasmonic Functionalized Metastructures to Enhance Local Heating Release}


\author{Antonio Ferraro}
\email[]{E-mail: antonio.ferraro@unical.it}
\affiliation{Physics Department, University of Calabria, I-87036 Arcavacata di Rende (CS), Italy}
\affiliation{CNR Nanotec-Institute of Nanotechnology, UOS Cosenza, 87036 Rende (CS), 87036, Italy}
\author{Giuseppe Emanuele Lio}
\affiliation{CNR Nanotec-Institute of Nanotechnology, UOS Cosenza, 87036 Rende (CS), 87036, Italy}
\affiliation{Physics Department, University of Calabria, I-87036 Arcavacata di Rende (CS), Italy}
\author{Abdelhamid Hmina}
\affiliation{Laboratoire Lumière, nanomatériaux \& nanotechnologies – L2n, Université de Technologie de Troyes \& CNRS ERL 7004, 12 rue Marie Curie, 10000 Troyes, France}
\author{Giovanna Palermo}
\affiliation{Physics Department, University of Calabria, I-87036 Arcavacata di Rende (CS), Italy}
\affiliation{CNR Nanotec-Institute of Nanotechnology, UOS Cosenza, 87036 Rende (CS), 87036, Italy}
\author{Thomas Maurer}
\email[]{E-mail:thomas.maurer@utt.fr}
\affiliation{Laboratoire Lumière, nanomatériaux \& nanotechnologies – L2n, Université de Technologie de Troyes \& CNRS ERL 7004, 12 rue Marie Curie, 10000 Troyes, France}
\author{Roberto Caputo}
\email[]{E-mail: roberto.caputo@unical.it}
\affiliation{Physics Department, University of Calabria, I-87036 Arcavacata di Rende (CS), Italy}
\affiliation{CNR Nanotec-Institute of Nanotechnology, UOS Cosenza, 87036 Rende (CS), 87036, Italy}
\affiliation{Institute of Fundamental and Frontier Sciences, University of Electronic Science and Technology of China, Chengdu 610054}

	
\begin{abstract}
	Plasmonic nanoheaters are reported that produce a significant local heating when excited by a $532\:nm$ wavelength focussed laser beam. A significant temperature increase derives from the strong confinement of electric field enabled by the specific arrangement of Au nanodisks constituting the nanoheater. The thermal response is much more sensitive when layering the gold nanoheaters by a thick layer of doped polymer, reaching a temperature variation of more than $250^\circ C$. The modulation of the excitation by a chopper enables the fine control of the thermal response with a measured maximum temperature variation of about $60^\circ C$ in a single period. These intriguing features can be efficiently exploited for the design of novel systems finding application in nano medicine and nano chemistry.
\end{abstract}


\keywords{Thermoplasmonic, nano heater, dye-doped, hybrid}

\maketitle 

\section{Introduction}
For many years, ohmic losses and related plasmonic heating generated by metallic micro- and nano-structures have been long seen as an issue, limiting the performance and functionalities of realized devices. However, in the last decade  many efforts have been spent by the research community towards the control of the thermal response of plasmonic structures for the applications in several fields like renewable energy, photo-catalysis and nano medicine, just to name a few \cite{boriskina2013plasmonic, fujimoto2014high, brongersma2015plasmon, besteiro2018, convertino2018array, roy2020temperature, baffou2020applications, kuppe2020hot, paria2020silver}. This significant interest arises from the feasible fine-control of the released heat by using lasers for the plasmonic excitation. The heat generation can be attributed to the resonant oscillation of free electrons, confined in metal nanoparticles (NPs), induced by the electric field carried on by the illuminating light source. The energy absorbed by the metallic system during this localized surface plasmon resonance (LSPR) excitation is then dissipated as heat into the surrounding medium following several phenomena including electro-electron scattering, electron-phonon and phonon-phonon coupling \cite{govorov2007generating, richardson2009experimental, baffou2010thermoplasmonics, govorov2014photogeneration, zhang2014optical, baffou_2017}. Key parameters for describing the performance of a thermoplasmonic system are absorption ($\sigma_{abs}$) and scattering ($\sigma_{sca}$) cross sections that measure how much light is absorbed and re-emitted respectively. The derivation of these quantities can be done by considering a metal sphere with diameter much smaller than the incident wavelength as an electromagnetic dipole \cite{baffou2013thermo}. In this approximation, the polarizability ($\alpha$) of such a sphere is defined as: 
\begin{equation}\label{eq:1}
	\alpha(\omega)= 4\pi R^{3}\frac{\epsilon(\omega)- \epsilon_{s}}{\epsilon(\omega)+ 2\epsilon_{s}}
\end{equation}
where R is the sphere radius, whereas $\epsilon(\omega)$ is the frequency-dependent complex permittivity of the NP immersed in a surrounding medium with $\epsilon(s)$ relative permittivity. Eq. \ref{eq:1} reveals that the resonance condition occurs when $\epsilon(\omega) \approx -2 \epsilon(s)$. 
At the resonance condition, ($\sigma_{abs}$) and ($\sigma_{sca}$) are then written as \cite{bohren1998absorption}:
\begin{equation}\label{eq:2}	
	\sigma_{abs}= kIm(\alpha) - \frac{k^{4}}{6 \pi}|\alpha|^{2}
\end{equation}
\begin{equation}\label{eq:3}	
	\sigma_{scat}= \frac{k^{4}}{6 \pi}|\alpha|^{2}
\end{equation}
where $k$ is the wave vector.
The extinction cross section ($\sigma_{ext}$) is instead calculated as:
\begin{equation}\label{eq:4}	
	\sigma_{ext}= \sigma_{abs}+ \sigma_{scat}= kIm(\alpha)
\end{equation}
LSPR wavelength and absorption cross sections are influenced by several factors as shape and material of the single NP, and spatial arrangement, in case of more particles \cite{mahi2017depth, marae2017angular, jauffred2019plasmonic}. In the last years, many attempt have been performed to actively control the of the absorption cross section of plasmonic systems including the use of flexible substrates \cite{cataldi2014growing, coppens2013probing, palermo2018flexible, burel2017plasmonic}, by introducing defects \cite{Liounclonable2020} or modifying  the surrounding medium \cite{pezzi2015photo, maurer2015beginnings, howard2017thermoplasmonic, roper2018effects, lio2020gamut}. \\
When considering the properties of a thermoplasmonic system, a key parameter is how generated heat is transferred to the environment. In heat diffusion, the thermal conductivity ($\kappa$) of the medium surrounding the plasmonic system becomes the main actor \cite{baffou_2017}. The heat transfer equation modeling the thermal behavior of the whole system is a function of the position dependent absolute temperature $T (\textbf{r})$ \cite{landau1987}: 
\begin{equation}\label{eq:5}	
	\rho C_{p} \frac{{\delta}T (\textbf{r})}{{\delta}t} = \nabla \cdot [\kappa \nabla T(\textbf{r}))] +Q 
\end{equation}
where $\rho$ and $C_{p}$ indicate density and specific heat capacity at constant pressure,  and $Q$ is the heat produced by an excited NP acting as a nano-source.
From the above formula, it is evident that the performance of plasmonic nano heater can be improved by appropriately choosing the surrounding medium. 
In this work, we present a detailed numerical and experimental study of the heat released by gold nano disks arranged in an optimized "flower" geometry. The overall size of an array of plasmonic flowers, patterned on glass substrate, is only $75 \times 75\:\mu m^2$. The investigated tiny area of the nanoheaters reveals a significant far field temperature variation $(\Delta T$, with respect to room temperature) from $\approx$ $30$ to $110^\circ C$, depending on the $532 nm$ incident laser intensity. The system performance is significantly enhanced by layering the nanoheaters with a $4\:\mu m$ thick Polyvinylpyrrolidone (PVP) film. In fact, due to the modification of the medium surrounding the nanostructures from air to PVP, the plasmonic coupling between close nanodisks is enhanced leading to an increase of the plasmonic coupling between neighbor nano flowers \cite{li2019large}. In this case, the generated $\Delta T$ reaches more than $200^\circ C$ in less than $1$ second. As predicted by Eq. 1, acting on the surrounding medium passively modifies the nanoheaters thermal response. A way to actively control the absorption cross section (and thus the system thermal response) is instead enabled by dissolving dye molecules (Rhodamine 6G) with maximum absorption at $\lambda=532 nm$ in the PVP mixture. When this second mixture is deposited on bare flower nanoheaters, even higher temperatures are reached upon excitation, until the photobleaching of the dye occurs. 
The dynamic response of the nanoheaters is experimentally investigated as well by periodically shutting the pump beam. The measured temperature profiles reveal a sensitive thermal response following the pump beam behavior up to 10 Hz frequency, with a $\Delta T$ excursion of more than $60^\circ C$ in one period. The results reported in this manuscript open a framework of possible applications where the fine control of heat generation is an essential parameter like in life science and nanotechnology. 
\section{Results and discussion}
\begin{figure}[!t]
	\includegraphics[width=1\columnwidth]{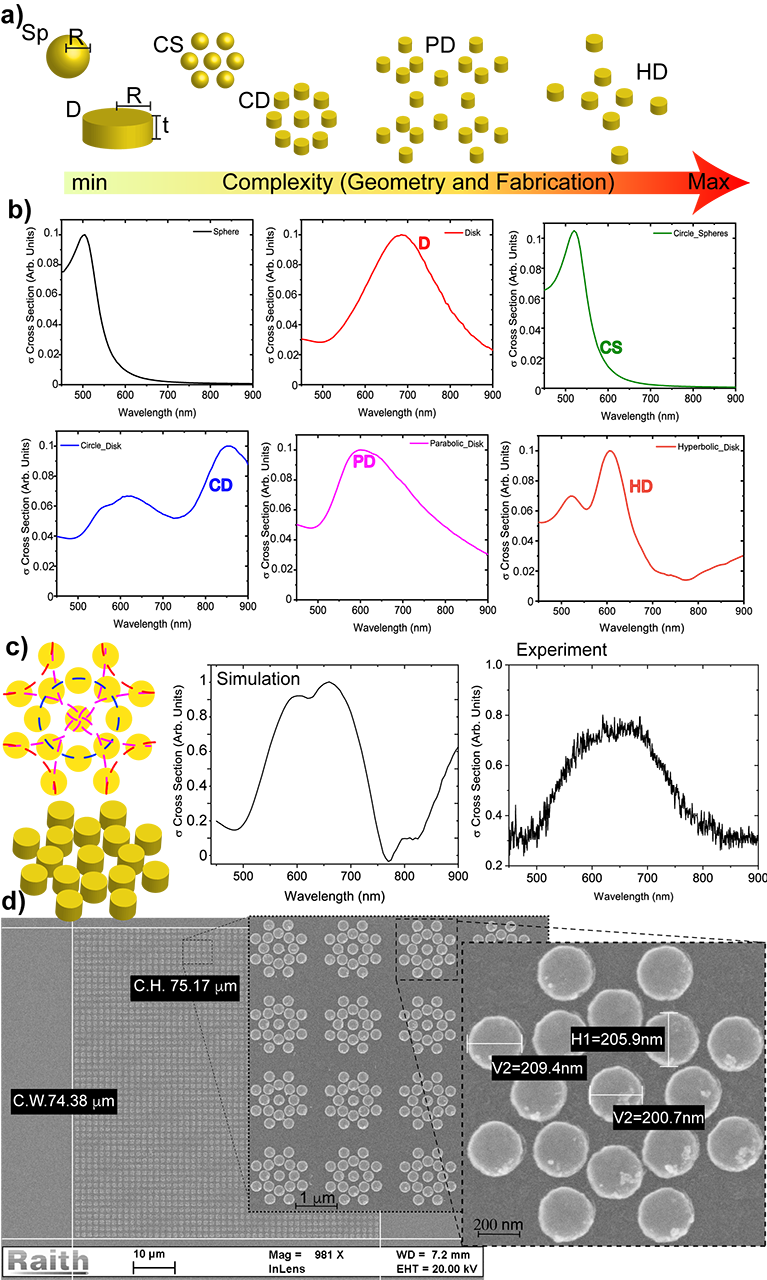}
	\caption{a) Schematic view of different nanostructures arrangement and b) their Scattering cross section c) Schematic view, Numerical and experimental cross section as a function of wavelength for the fabricated nano flowers d) Scanning Electron Microscope (SEM) micro-graphs of the fabricated plasmonic nano heater in flower arrangement.}
	\label{1}
\end{figure}
\begin{figure}[!h]
	\includegraphics[width=1\columnwidth]{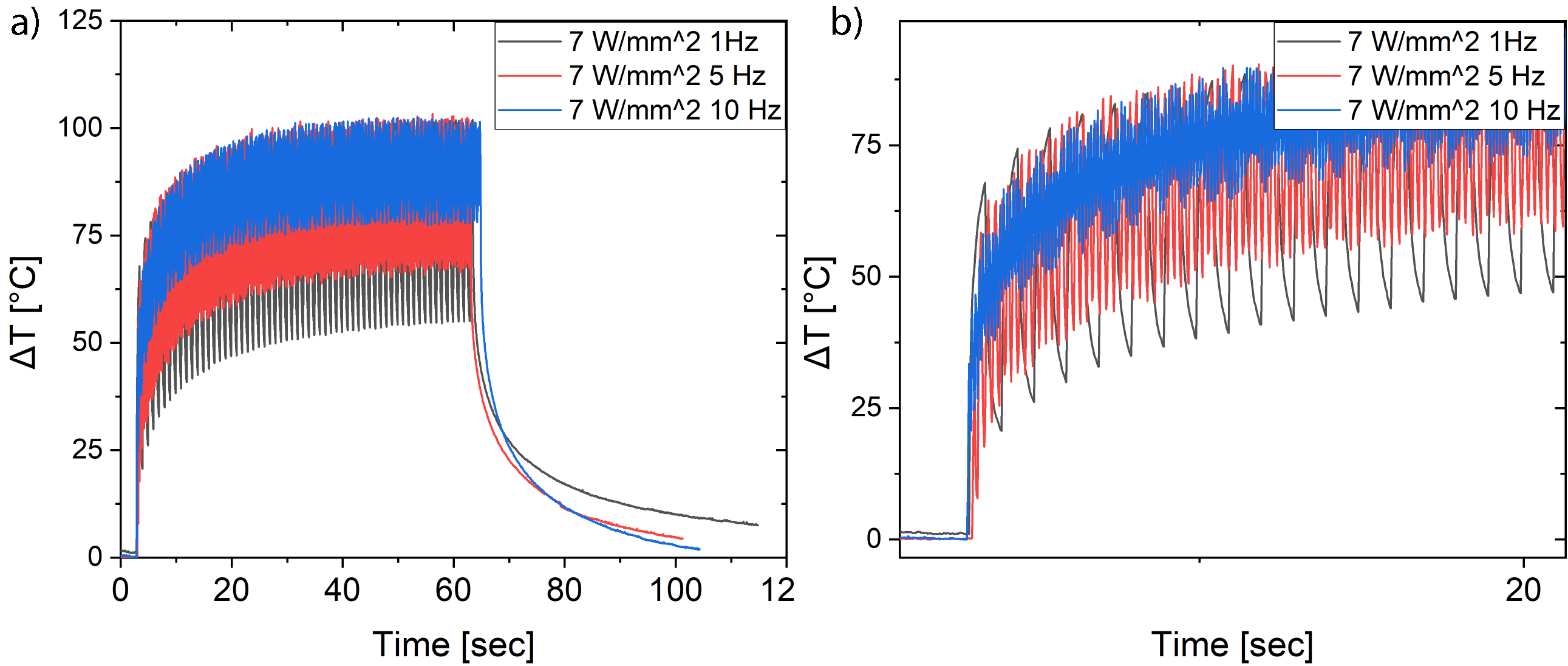}
	\caption{ a)Comparison of experimental far-field temperature acquired using an intensity of $7 W/mm^{2}$ and laser chopper frequency of $1 Hz$, $5 Hz$ and $10 Hz$ for gold nano flower covered with $4 \mu m$ thick PVP +R6G layer b) zoom of the measurements.}
	\label{6}
\end{figure}

The search for the optimal distribution of metal nanoparticles or nanostructures, able to maximize the heat generation, represents an open challenge. 
The basic idea is to shift the absorption cross section ($\sigma_{abs}$) so as to match the laser pump line, thus increasing the thermal variation. To this aim, several attempts have been made by optimizing the NPs distribution thus creating efficient hot-spots and square, diagonal, octagonal and circular patterns have been considered \cite{kudyshev2019machine, lio2019opto,lio2019tensile, kudyshev2020machine, ashalley2020multitask}. As an approach to challenge the problem in a more systematic way, we implemented a Matlab code that considers main analytical curves as circles, parabolas and hyperbolas as traces along which Au spheres or disks can be positioned (Figure \ref{1}a). For each obtained arrangement, COMSOL Multiphysics calculates the corresponding absorption cross section that is strongly dependent on the given design in the investigated spectral range (Figure \ref{1}b). The absorption values of all calculated curves have been normalized between zero and one to simplify their comparison, more details in \textit{Experimental section}. By combining the previous spatial distributions and considering the steric hindrance of the particles, pump wavelength $\lambda = 532\:nm$, polarization insensitiveness, broad $\sigma_{abs}$ and, at least, $\Delta T>15^\circ C$, the optimal arrangement is found that resembles the "flower" illustrated in Figure \ref{1}c. This pattern comprises disks with $R\sim100\:nm$ and $t\sim50\:nm$ and exhibits a broad numerical cross section that is in good agreement with the corresponding experimental one (Figure \ref{1}c). The broad absorption spectrum, from $500 nm$ to $750 nm$, reveals quite convenient in case the system is excited by multiple wavelengths or white light. For what concerns the absolute values of absorption, the flower configuration, with the same number of disks of the other arrangements (CD, PD, and HD), exhibits a $\sigma_{abs}$ at least one order of magnitude larger.
After design and optimization process, the Au nanoheaters were fabricated through Electron Beam Lithography (EBL) on glass substrate, see Experimental Section. However, the configuration of this arrangement with specific gaps between disks needs an investigation of the electron beam exposure dose to obtain the best resolution of the openings created in the PMMA resist. In fact, the dose factor depends on the pattern shape and especially on the gaps range ($10 nm$ to $50 nm$). A dose variation in the range $200 - 400\:\mu C/cm^2$ was applied to adjust the EBL process. Finally two different doses have been used: center and external disks with $340\:\mu C/cm^2$ while $280\:\mu C/cm^2$ for inner disks where the gap sizes are smaller and the local proximity effect caused by the forward scattering electrons is high. The final device results in an array of flower nanoheaters with a period of $1.5 \mu m$ for a total patterned area of $75 \times 75\:\mu m^2$. In the single flower, the central disk is positioned at a distance of $\approx 100\:nm$ from the first neighbors, whereas the other disks are spaced of $\approx 40-50\:nm$.
The morphology of the fabricated nano flowers has been investigated by scanning electron microscopy (SEM) and the acquired images are reported in Figure \ref{1}d where in the first frame the $75 \times 75\:\mu m^2$ array is shown that contains about 5000 flowers, in the second one ($5\:\mu m \times 5\:\mu m$), the detail of the ordered arrangement of these flowers is reported together with a single zoomed-in flower; the third frame displays the disk diameter and high resolution morphology details. All frames underline the excellent quality of the realized fabrication process. \\
\begin{figure}
	\includegraphics[width=1\columnwidth]{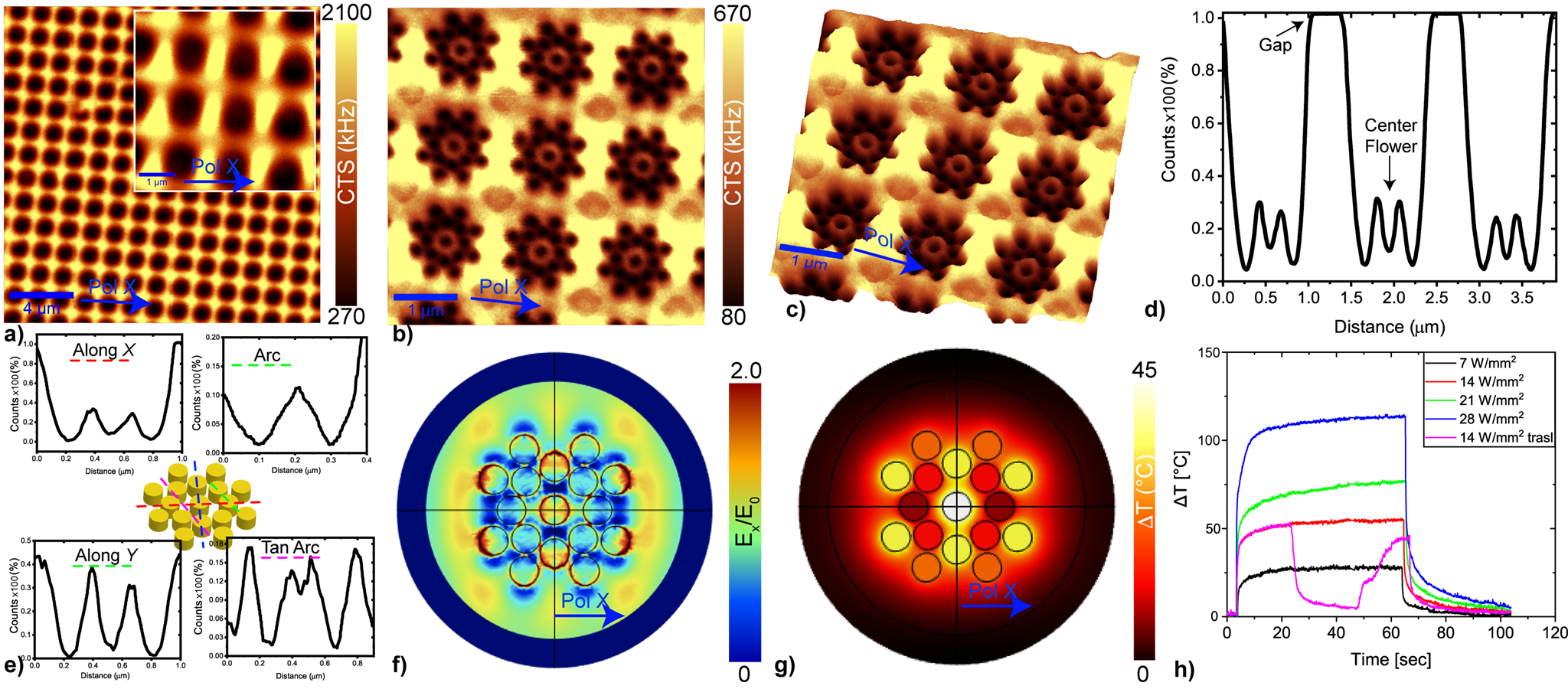}
	\caption{ a) Experimental confocal images of the plasmonic nano heater  where the black zone denotes the nano flower. b) Experimental Scanning Near-Field  Optical Microscope (SNOM) of the plasmonic nano heater $5\times5 \:\mu m$ where the black zone denotes the Au nano flower, c) 2.5D view of micrograph (b). d) Near Electric-Field profile on the averaged area $5\times5\:\mu m$ and e) near field distribution  profiles along disks at different positions. f) Normalized Numerical Electric Field Map and g) Near field Thermal Map of the proposed geometry using an intensity of $14 W/mm^2$, h) Experimental far-field temperature at different laser ($\lambda =532nm$) intensity for $60 sec$ of irradiation. The purple line refers to a measurement acquired moving, after $20 sec$, the nano flower far away from the laser spot.}
	\label{2}
\end{figure}
In order to study the plasmonic behavior of the flower nanoheaters upon excitation, confocal and scanning near-field optical microscope (SNOM) analyses have been performed. As shown in Figure \ref{2}a, the far-field maps acquired in transmission by the bottom objective lens ($60\times$) of the confocal microscope evidence a prevalent field enhancement at the external borders of each nano-flower, with values in the inter-flower gap $\approx$ 7 times higher than those measured within the single structure, see inset for more details. Light is strongly absorbed by the flowers that appear as black spots without a possibility to discern the fine nanodisk morphology. Higher resolution SNOM images (Figure \ref{2}b,c) evidence more detailed field maps highlighting the inner contours of the beautiful flowers. The near field images also provide a meaningful insight of their collective plasmonic response \cite{shimada2013spatial, zhou2015two, bazylewski2017review, wang2020three}. In particular, the nanodisks belonging to a single flower are strongly coupled between each other because of their extreme proximity and all together behave as a single plasmonic macro-unit. Overall, the flower undergoes a dipolar excitation whose field lobes are oriented along the exciting polarization direction (x). A donut-shaped field localization is observed around the central disk, probably due to its distancing from first neighbors ($\approx 100\:nm$), that weakens the plasmonic coupling. However, the several transverse field cuts reported in Figure \ref{2}d,e show that the donut field amplitude is quite moderate if compared to that of the outer lobes. The observed plasmonic behavior is in significant agreement with the numerical simulation of a single flower which confirms the presence of dipolar field lobes as a result of an excitation along the x direction (Figure \ref{2}f). In the center of the simulated flower, an elongated field hot-spot is instead present that is in contrast with the experimentally observed donut shape. This discrepancy can be probably attributed to the slightly different excitation conditions between the experimental and numerical case. While in the numerical case an ideal plane wave is assumed for the excitation, in the experimental one the exciting light is focussed on the sample by passing through the SNOM tip with a hole radius of $\approx 90\:nm$. Considering that the aperture size is comparable with the interdistance between central disk and neighboring ones, the measurement is possibly affected by a modification of the light polarization that results in the observed centro-symmetric donut field pattern. The numerically simulated thermal response of the flower nanoheater is performed by utilizing the same excitation conditions considered for the near-field simulation (Figure \ref{2}f). The confinement and enhancement of electric field yields an increase of temperature around the Au nano flower (Figure \ref{2}g). It is worth noting a $\Delta T = 46^\circ C$ in the central hot-spot, due to the plasmonic coupling with the neighboring Au nanodisks, during light excitation. Details about the performed numerical simulation are available in the Experimental section. \\
Following the optical characterizatiom of the device, its thermal response has been experimentally investigated by utilizing a IR thermocamera that provides a far-field information (Figure \ref{2}h). For each measurement, the sample has been irradiated for $60 s$ by a $\lambda = 532 nm$ laser set at different intensities in the range $7$-$28 W/mm^2$, see Experimental section for details. 
The nano heaters exhibit a very fast response reaching a temperature equilibrium in less then 10 second of excitation with a maximum $\Delta T$ passing from $28^\circ C$ to $110^\circ C$, depending on the beam intensity. It is worth nothing that for $14 W/mm^2$ the experimental $\Delta T$ (red curve) is almost equal to the numerical one reported in Figure \ref{2}g. Since the patterned area is quite small ($75 \times 75\:\mu m^2$), it is expected that a slight movement of the nanoheaters from the laser spot results in a drastic temperature decrease. This is experimentally confirmed by the purple curve in Figure \ref{2}h where, after $20 s$ of excitation, the sample was moved away by hundred microns and then back again.
\begin{figure}[!t]
	\includegraphics[width=1\columnwidth]{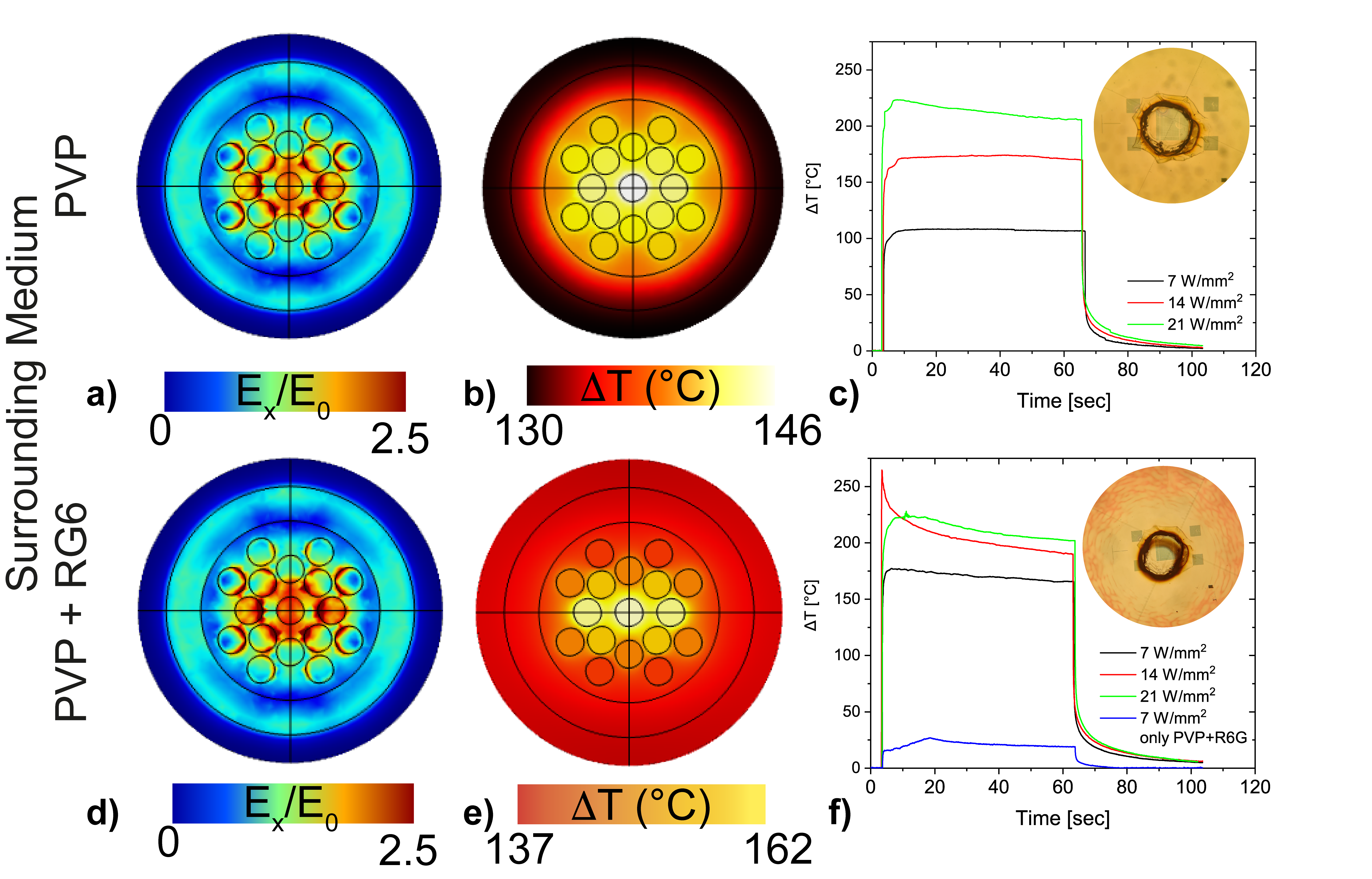}
	\caption{ a) Normalized Numerical Electric Field Map and b) Near field Thermal Map of the proposed geometry embedded into $4 \mu m$  thick PVP layer,c)Experimental far-field temperature at different laser ($\lambda =532nm$) intensity for $60 s$ of irradiation. Inset: optical micrograph of the nanodisks where the black circle represent dissolved/burnt PVP  d) Normalized Numerical Electric Field Map and e) Near field thermal map of the proposed geometry embedded into  $4 \mu m$ thick PVP+R6G dye layer,f) Experimental far-field temperature at different laser ($\lambda =532nm$) intensity for $60 s$ of irradiation. The blue line refers to a measurement where only PVP+R6G are present. Inset: optical micrograph of the nano flower where the black circle represent dissolved/burnt PVP +R6G}
	\label{3}
\end{figure}
For enhancing their thermal emission, the nano heaters have been completely embedded in a matrix of $4 \mu m$ thick of Polyvinylpyrrolidone (PVP) characterized by a thermal conductivity ($\kappa$) higher than that of air. 
As predicted by Eq. \ref{eq:1}, the change of dielectric constant from $1$ to $\sim2.03$ sensibly modifies the plasmonic response of the system with an overall increase of the normalized electric field and generation of a larger hot-spot cluster in the center of the nano flower (Figure \ref{3}a). The numerical results also confirm the more efficient thermal behavior: a pump intensity of $14 W/mm^2$ results in a more sensitive heat generation with $\Delta T = 146^\circ C$ in the hot-spot that decreases to $\approx 130^\circ C$ at the boundary of the nano flower (Figure \ref{3}b). This high temperature variation is confirmed by the experimental measurements, red curve in Figure \ref{3}c. Moreover, by irradiating with the minimum intensity of $7 W/mm^2$ the $\Delta T = 100^\circ C$ is easily reached, while the maximum intensity used in this work, namely $21 W/mm^2$, increases $\Delta T > 200^\circ C$. This extremely high confinement of the electric field and hence of the photogenerated heat is confirmed by the optical image, acquired after measurement and reported in the inset of Figure \ref{3}c that highlights a circle of burned PVP around the nano flower matrix.
The thermal emission is strongly related to the absorption cross section of the considered system. The addition of specific fluorescent molecules with an absorption band that overlaps the absorption cross section represents an easy way to enhance the whole absorption cross section and hence the thermal emission. The rhodamine 6G (R6G) Dye, with a maximum absorption exactly at the same wavelength of the pump laser source ($\lambda=532\: nm$), represents the ideal candidate to maximize the emission of the proposed device.   
The polymeric material doped with the dye emitter (PVP+ Rhodamine 6G) can be modeled by considering a combination of Lorentzian and Gaussian functions \cite{ fofang2011plexciton}. The expression for the complex dielectric constant of the doped polymer $\tilde{\varepsilon}$, taking into account the dye broadband absorption and emission, is written as:
\begin{eqnarray}\label{eq:6}
		\tilde{\varepsilon}=\varepsilon_{polymer} fs_{abs}\frac{\omega_a^2}{\omega_a^2-\omega_0^2-i\gamma_a\omega}(n^a_0-n^a_1) + \nonumber\\ \varepsilon_{polymer} fs_{abs}exp\left [ \frac{(\omega-\omega_a)^2}{2(-i\gamma_a2\omega)^2}\right ] \nonumber\\ - \varepsilon_{polymer} fs_{em}\frac{\omega_e^2}{\omega_e^2-\omega_0^2-i\gamma_e\omega}(n^e_0-n^e_1) \nonumber\\ - \varepsilon_{polymer} fs_{emi}exp\left [ \frac{(\omega-\omega_e)^2}{2(-i\gamma_e2\omega)^2}\right ]
	\end{eqnarray}
where $\omega_a$, $fs_{abs}$ and $\gamma_a$ are angular frequency, strength and width of absorption, respectively. The emission, modeled by a Lorentzian function, is characterized by a strength parameter $fs_{em}$, an angular frequency $\omega_e$ and a curve width $\gamma_e$.  The parameter $n^{a(e)}_m$ represents the occupied energetic states in a two level model, with the populations $n_0$ of the ground state and $n_1$ of the excited state. In this model, the R6G dye is characterized by an absorption peak at $\sim 530 \:nm$ and an emission peak at $\sim615\: nm$. \\
With respect to the case with PVP only, the behavior reported in Figure \ref{3}d,e is similar but characterized by higher temperature, with a net increase of about $70^\circ C$ for the lowest pump intensity black curve in Figure \ref{3}f). However, in presence of the dye, the underlying physical phenomena taking place are different. In fact, an extremely interesting behavior occurs with an intensity of  $14 W/mm^2$ (red curve): a very fast temperature increase of $264^\circ C$ in $500 ms$ is followed by an immediate drastic decrease. This fast decay is probably due to the irreversible photobleaching of the dye molecules that accompanies the well observable "burning" of the PVP (inset of Figure \ref{3}f). A confirmation of the irreversibility of the dye photo-bleaching is obtained by pumping the same area of the sample with a higher intensity, $21 W/mm^2$ (red curve). The result of the thermal measurement is now analogue to the one acquired without dye, indicating that, being the dye already unable to fluoresce, the system behaves as it is not present at all. As observed in the previous case (inset of Figure \ref{3}c), the large amount of released heat burns the polymer only in the area surrounding the nano heaters. By moving the sample away from the nano-flower, the PVP+dye layer produces a $\Delta T$ of only $25^\circ C$ (blue curve in Figure \ref{3}f).\\
\begin{figure}[!h]
	\includegraphics[width=1\columnwidth]{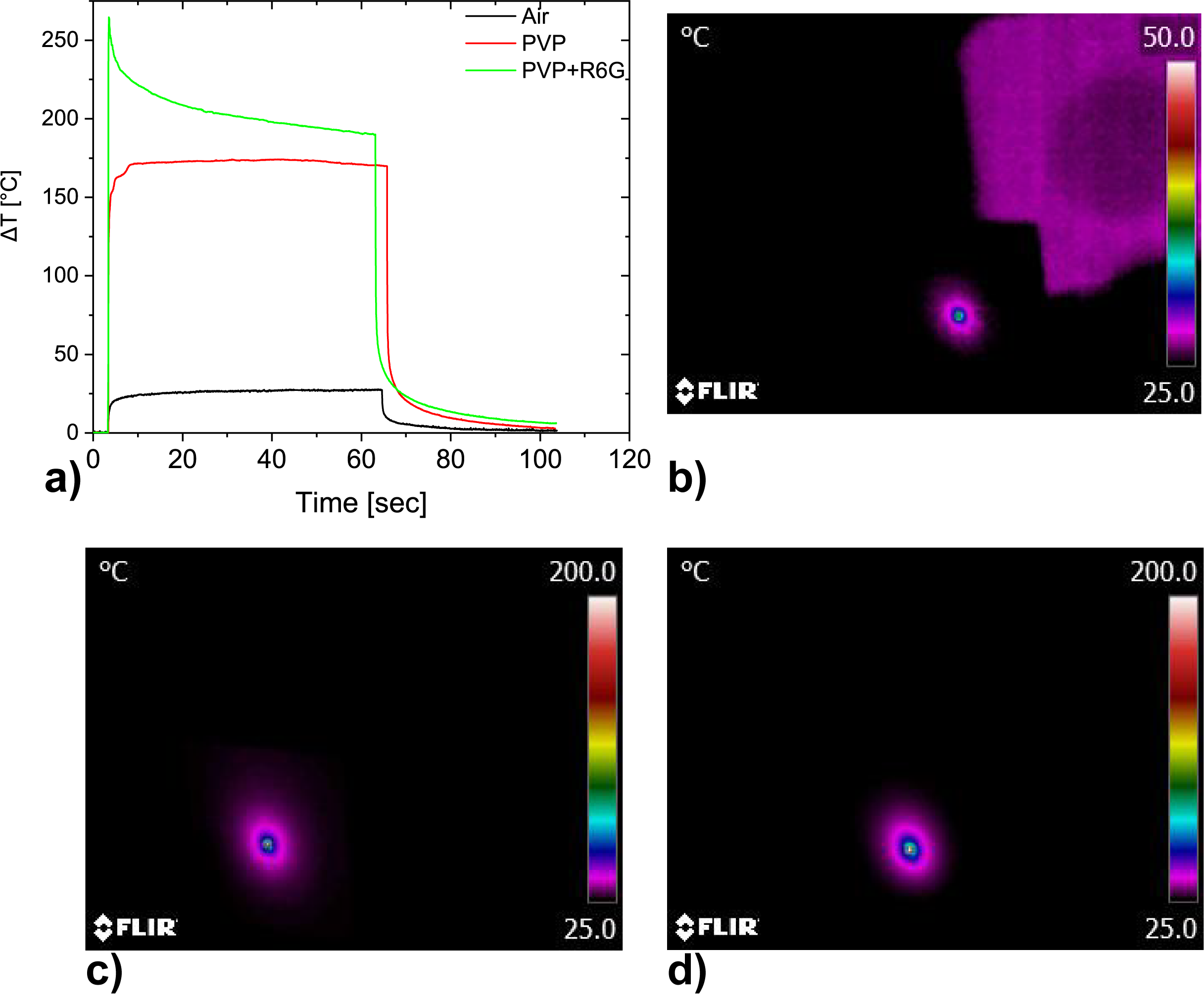}
	\caption{ a) Comparison of experimental far-field temperature acquired using an intensity of $14 W/mm^{2}$.  Experimental far-field thermal map after 30" of excitation of b) nano flower, c) nano flower covered with $4 \mu m$ thick PVP layer, d) nano flower covered with $4 \mu m$ thick PVP +R6G layer}
	\label{4}
\end{figure}
A rapid overview of the thermal response of the three configurations (air, PVP and PVP+dye) at a fixed laser intensity of $14 W/mm^2$ is illustrated in Figure \ref{4}a. The $\Delta T$ enhancement of the nano flower covered by PVP+R6G is clearly visible with respect to the other cases. The maximum $\Delta T$ is higher than $250^\circ C$ but, due to bleaching of the dye, this value rapidly decreases to about $200^\circ C$. The far field thermal maps after $30$ seconds of excitation are reported in Figure \ref{4}b-d, highlighting the highest temperature positions where the laser beam exactly impinges on the sample.
\begin{figure}
	\includegraphics[width=1\columnwidth]{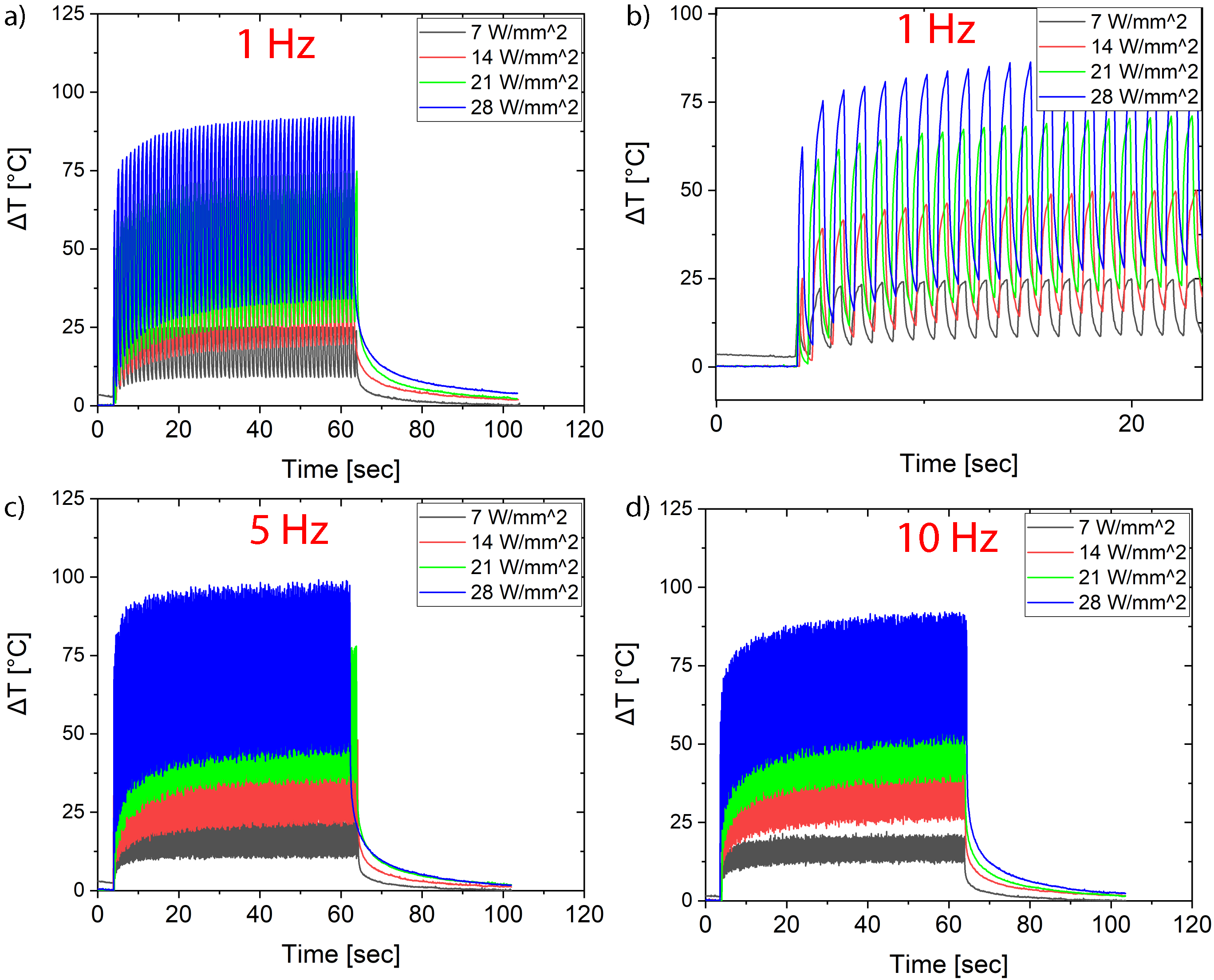}
	\caption{ Experimental far-field temperature of gold nano flower at different laser ($\lambda =532nm$) intensity and chopper frequency of a) $1 Hz$, b) zoom of $1 Hz$, c) $5 Hz$ and d) $10 Hz$.}
	\label{5}
\end{figure}
The dynamic properties of the nano heaters thermal response have been investigated by pumping the bare system (nano flowers exposed to air) at the previously utilized laser intensities while periodically shutting the laser beam with a chopper rotating at three different frequencies, namely $1 Hz$, $5 Hz$ and $10 Hz$. This test provides information about how fast the system is able to dissipate the photo-generated heat. The results (Fig. \ref{5}) reveal a lower $\Delta T$ than the one measured in the corresponding static cases. This suggests that the time needed to generate all the possible heat is longer than the one allowed by the chopper rotation. For all the considered chopper frequencies, it is demonstrated that the $\Delta T$ value achieved in a period is directly proportional to the pump intensity and inversely proportional to the chopper frequency, with an optimal result of $\approx$ $60^\circ C$ for the case of $ 28 W/mm^{2}$ and $1 Hz$ (Fig. \ref{5}b). 
The same dynamic measurements have been performed on the nano flowers covered by PVP+R6G. For all chopper frequencies, a pump intensity of $7 W/mm^{2}$, well below the photo-bleaching threshold of $14 W/mm^{2}$, has been used (Fig. \ref{6}). As expected, the presence of the R6G dye enhances the thermal response of the investigated system thus resulting in a wider temperature variation in a period, almost double with respect to the bare nano flower case.  
\section{Conclusion}
This manuscript presents a new thermoplasmonic device composed of gold nano heaters, operating at wavelength of $532 \:nm$, showing an extremely high $ \Delta T$ of more than $100^\circ C$. Numerical and experimental measurements have been performed on bare structures showing the expected characteristics such as broad absorption cross section and high thermal heating. The performance of the proposed nanostructure has been improved by covering the nano heaters with a layer of polymer (PVP) enabling a temperature increment of almost $210^\circ C$. In fact, the presence of a cover layer strongly enhance the confinement of the electric field. Successively, a dye, absorbing at the operating wavelength, has been added to the PVP solution and spin coated on the bare device. The augmented absorption cross section, now, permits to reach $ \Delta T$ higher than $250^\circ C$ in less than 1 second. \\
Finally, by using a chopper to regularly interrupt the impinging laser beam, a new temperature control mechanism has been added to thermoplasmonic nanostructures. In this situation the maximum temperature variation reached in a period of $1Hz$ is $\approx$ $60^\circ C$.\\ 
To conclude, the achieved results can pave the way for new breakthrough applications in several field that cover a paramount interest in chemistry, life science and renovable energy.


\section{Experimental Section}
\medskip
\textbf{Experimental Section} \par
\textbf{Numerical Model}: The absorption and scattering cross sections have been calculated using a Finite Element Method (FEM) implemented in COMSOL Multiphysics by means of the following relations: $\sigma_{abs}$=$W_{abs}/P_{in}$, $\sigma_{sca}$=$W_{sca}/P_{in}$, where $P_{in}$ is the incident irradiance, defined as energy flux of the incident wave; $W_{abs}$ is the energy rate absorbed by particle, that is derived by integrating the energy loss $Q_{loss}$ over the volume of the particle, while $W_{sca}$ is the energy rate absorbed by particle, is derived by integration of the Poynting vector over an imaginary sphere around the particle.
All the other parameters as scattered field method, proper polarization and propagation direction as opportunely set to numerically investigate the near field distribution and the plasmonic response of the nanodisks. The light source used in the FEM software is a monochromatic one, meaning that the spectral simulations are performed wavelength by wavelength.
The light beam intensity has been evaluated as $I= E_0^{2}/(2 Z_{0const})$ = 13.27 W/mm$^{2}$, where $E_0$ is the initial electric field ($1\times 10^{5}\:V/m$) and $Z_{0const} $= 376.73 $\Omega$ is the impedance of the system that takes into account also the incident area. This tool package offers the possibility to study the energy flow that passes through the nanodisk when they are shone by a light beam and the energy flow scattered from nano flower and collected on the external layer used as an integrating sphere. \\
\textbf{Nanofabrication}:
The nano flowers have been fabricated using Electron Beam Lithography (EBL) Raith system operating at $20 kV$ and $10\: \mu m$ aperture to reduce the neighboring effects. A an e-beam resist, polymethyl methacrylate PMMA 950 k (molecular weight) was spin coated at $4000 rpm$ for $30 seconds$ producing a layer of $160\: nm$. The glass sample was baked at 170 °C for 15 minutes. A thin layer of conductor polymer was spin coated to provide anti-charging during the adjustment of focus and astigmatism by a contamination spot observed at fixed point exposure of substrate top surface. The exposure of PMMA layer resist was carried out with a spot size of a $3\: nm$ and a beam current of $10 pA$. Next, the development was performed in a (1:3) MethylIsoButylKetone (MIBK): Iso-propyl alcohol (IPA) solution for $60 s$ at $23^\circ C$ and finally  rinsed in IPA during $30 seconds$, dried with air and followed by electron beam deposition process of $5\: nm$ Cr layer as an adhesion promotor. A $50\: nm$ gold layer was deposited with deposition rate of $0.1\: nm/s$. Finally the sample was immersed in acetone to remove the PMMA layer and the unwanted gold layer (lift –off).\\
\textbf{Confocal and Scanning Near-Field Optical Microscope}:
The confocal images have been acquired using  WITEC Alpha 300, in transmission mode by using as source a laser having  wavelength of $532 nm$ X-polarized focused on the sample through a Zeiss 100x objective confocal with a Leica 60x objective, used at the bottom for collection. A PMT (Photomultiplier tubes) has been used for the detection. 
For SNOM measurement, the same instrument has been used but it is equipped with hollow AFM tips covered with aluminium. Then, an incident wavelength of $532 nm$ has been focused on the sample passing through the SNOM tip with a hole radius of $90\:nm$\\
\textbf{Polymer solution} A Polyvinylpyrrolidone (PVP) solution, dissolved in ethanol with a final concentration of $15 wt\%$, has been spin coated at 2000rpm for 30 seconds followed by a baking on hot plate at $80^\circ C$ for 5 minutes.
A solution of Rhodamine 6G $5 wt\%$ in ethanol has been prepared and mixed in a ratio 1:1 with the PVP $15 wt\%$ solution. The PVP+R6G solution has been spin coated coated at 2000rpm for 30 seconds followed by a baking on hot plate at $80^\circ C$ for 5 minutes.
\textbf{Optical Characterization}: The experimental cross-section has been acquired using an optical microscope in transmission mode with 50x objective for acquiring the light only from the area where the flower arrangment is fabricated. . The light is collected with an optical fiber connected to UV-Vis spectrometer Ocean Optics USB2000+. \\
\textbf{Thermal Measurement}:
For far-field thermal measurement, a COHERENT Verdi $\lambda=532nm$ CW laser has been used as source. A pin-hole with a diameter of 3mm has been used to reduce the beam spot. Successively, the laser beam has been focused on the sample through a lens producing a spot of around $100 \mu m$. The far-field thermal measurement have been acquired using a IR camera FLIR E40 having an IR resolution of 160x120 pixel and a spatial resolution (IFOV) of $2.72 mrad/pixel$. The emissivity has been fixed at 0.97 which was experimentally retrieved using the relative procedure.
The measurement start 3 seconds prior the beam impinges on the sample for 60 seconds. Then, the beam is shutted-off and the temperature decay is measured 40 seconds. \\

\section*{Acknowledgements} 
A.F. G.P and R.C thank the ``Area della Ricerca di Roma 2", Tor Vergata, for the access to the ICT Services (ARToV-CNR) for the use of the COMSOL Multiphysics Platform and Origin Lab, and the Infrastructure ``BeyondNano" (PONa3-00362) of CNR-Nanotec for the access to research instruments.\\
This work has been supported by the Agence Nationale de la Recherche and the FEDER (INSOMNIA project, contract "ANR-18-CE09-0003"). Financial support of NanoMat ( www.nanomat.eu) by the “Ministère de l’enseignement supérieur et de la recherche,” the “Conseil régional Champagne-Ardenne,” the “Fonds Européen de Développement Régional (FEDER) fund,” and the “Conseil général de l’Aube” is also acknowledged by A.H and T.M
\medskip

\section*{References}
\newpage

%

\end{document}